\documentclass[12pt]{article}
\usepackage{pdproc,epsfig,graphics} 
\begin{document}

\newcommand{\lsim}   {\mathrel{\mathop{\kern 0pt \rlap
  {\raise.2ex\hbox{$<$}}}
  \lower.9ex\hbox{\kern-.190em $\sim$}}}
\newcommand{\gsim}   {\mathrel{\mathop{\kern 0pt \rlap
  {\raise.2ex\hbox{$>$}}}
\lower.9ex\hbox{\kern-.190em $\sim$}}}
\def\be{\begin{equation}}
\def\ee{\end{equation}}
\def\ba{\begin{eqnarray}}
\def\ea{\end{eqnarray}}
\def\d{{\rm d}}
\def\ap{\approx}

\title{
SUPER-GZK NEUTRINOS}

\author{V. Berezinsky}

\address{INFN, Laboratori Nazionali del Gran Sasso,\\ 
67010 Assergi (AQ) Italy}

\abstract{ 
The sources and fluxes of superGZK neutrinos, $E>10^{20}$~eV, 
are discussed. The fluxes of {\em cosmogenic neutrinos}, i.e. those 
produced by 
ultra-high energy cosmic rays (UHECR) interacting with CMB photons, are
calculated in the models, which give the good fit to the observed flux of  
UHECR. The best fit given in no-evolutionary model with maximum
acceleration energy  $E_{\rm max}=1\times 10^{21}$~eV results in very
low flux of  superGZK neutrinos an order of magnitude lower than the observed 
flux of UHECR. The predicted neutrino flux becomes larger and
observable by next generation detectors at energies $10^{20} - 10^{21}$~eV
in the evolutionary models with $E_{\rm max}=1\times 10^{23}$~eV.
The largest cosmogenic neutrino flux is given in models with very flat
generation spectrum, e.g. $\propto E^{-2}$.
The neutrino energies are naturally high in the models of
{\em superheavy dark matter and topological defects}. Their fluxes can
also be higher than those of cosmogenic neutrinos. The largest fluxes are given
by  {\em mirror neutrinos}, oscillating into ordinary neutrinos. Their 
fluxes obey some theoretical upper limit which is very weak, and in practice 
these fluxes are most efficiently limited now by observations of radio
emission from neutrino-induced showers. 
}

\normalsize\baselineskip=15pt

\section{Introduction}
The abbreviation `SuperGZK neutrinos' implies neutrinos with energies
above the Greisen-Zatsepin-Kuzmin \cite{GZK} cutoff 
$E_{\rm GZK} \sim 5\times 10^{19}$~eV. Soon after theoretical
discovery of the GZK cutoff, it has been realized that this phenomenon is 
accompanied by a flux of UHE neutrinos, which in some models can be
very large \cite{BZ}. In 80s it was understood that 
topological defects can
produce unstable superheavy particles with
masses up to the  GUT scale \cite{Witten} and neutrinos with
tremendous energies  can emerge due to this process \cite{HiSch}. 

It has been proposed that SuperGZK 
neutrinos can be detected observing the horizontal Extensive Air Showers  
(EAS) \cite{BeSm}. 
The exciting prospects for detection of SuperGZK neutrinos have appeared 
with the ideas of space detection, e.g. in the projects 
EUSO\cite{EUSO} and OWL\cite{OWL}. The basic idea of detection can
be explained by example of EUSO. 

The superGZK neutrino entering the Earth atmosphere in near-horizontal 
direction produces an EAS. The known fraction of its energy, which
reaches 90\% , is radiated in form of isotropic fluorescent light. 
An optical telescope from a space observatory detects this
light. Since the observatory is located at very large height 
($\sim 400$~km) in comparison with thickness of the atmosphere, the
fraction of detected flux is known, and thus this is the calorimetric 
experiment (absorption of light in the upward direction is 
small). A telescope with diameter 2.5 m controls the area 
$\sim 10^5$~km$^2$ and has a threshold for EAS detection 
$E_{\rm th}\sim 1\times 10^{20}$~eV.     

The very efficient method of superGZK neutrino detection is given by 
observations of radio emission by neutrino-induced showers in ice, salt 
and lunar regolith.
This method has been originally suggested by G.~Askaryan 
in 60s \cite{askarian}. Propagating in the matter the shower acquires 
excessive negative electric charge due to involvement of the 
matter electrons in knock-on process. The coherent Cerenkov radiation
of these electrons produces the radio pulse. Recently this method has
been confirmed in the laboratory measurements \cite{saltz}. There were 
several searches for such radiation from
neutrino-induced showers in the Antarctic and Greenland ice and in the
lunar regolith. In all cases the radio-emission can be observed only 
for neutrinos of extremely high energies.
The upper limits on the flux of these neutrinos 
have been obtained: in GLUE experiment \cite{glue} by radiation from
the moon, in FORTE experiment \cite{forte} by radiation from the
Greenland ice and in RICE experiment \cite{rice} from the Antarctic ice.

The characteristic feature of the detection methods described above is 
the high energy threshold, typically 
$E \gsim 1\times 10^{19} - 1\times 10^{20}$~eV. How neutrinos of these 
energies can be produced?      

The most conservative mechanism of superGZK neutrino production is
$p\gamma$ mechanism of collisions of accelerated protons/nuclei with 
low-energy CMB photons. To provide neutrinos with energies higher that 
$1\times 10^{20}$~eV the accelerated protons must have energies higher 
(or much higher) than $2\times 10^{21}$~eV. For shock acceleration 
this energy can reach optimistically $1\times 10^{21}$~eV. One
has raise his hopes on less developed ideas of acceleration such as 
acceleration in strong e-m waves, exotic plasma mechanisms of
acceleration and unipolar induction. 

The top-down scenarios can easily provide neutrinos with energies
higher and much higher than $1\times 10^{20}$~eV. The idea common  
for many mechanisms is given by existence of superheavy particles with 
very large masses up to GUT scale. In Grand Unified Theories (GUT)
these particles (gauge bosons and higgses) are short-lived. In the
cosmic space they are produced by Topological Defects (TDs). The
decay of these particles results in the parton cascade, which is 
terminated by production of pions and other hadrons.
Neutrinos are produced in their decays.

The superheavy particles are naturally produced at post-inflationary 
stage of the universe. The most reliable mechanism of production 
is gravitational one. The masses of such particles can reach 
$10^{13} - 10^{14}$~GeV. Protecting by some symmetry (e.g. gauge
symmetry or discrete gauge symmetry like R-parity in supersymmetry),
these particles can survive until present cosmological epoch and
produce neutrinos in the decays or annihilation. 
 
\section{Upper limits on superGZK neutrino flux}
There are two different upper limits on UHE neutrino fluxes: cascade 
upper limit \cite{BeSm} and cosmic ray upper limits (Waxman-Bahcall 
\cite{WB} and Mannheim-Protheroe-Rachen \cite{MPR}). The cosmic ray
upper limits are not relevant for superGZK neutrinos because this limit
is not valid for top-down scenarios and it is automatically satisfied
for cosmogenic neutrinos, since their fluxes are calculated in the
models which explain the observed UHECR. 

\begin{figure*}[h]
  \begin{center}
  \mbox{\includegraphics[width=0.7\textwidth]{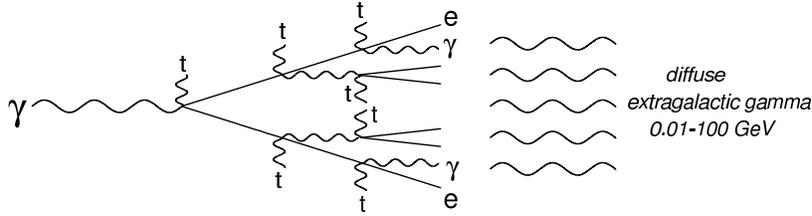}}
  \end{center}
  \caption{ Developing of e-m cascade in collisions with background target 
           (t) photons.}
\label{cascade}
  \end{figure*}

The cascade upper limit on HE and UHE neutrino fluxes \cite{BeSm,book}
is provided due to \mbox{e-m} cascades initiated by HE photons or electrons  
which always accompany production of neutrinos. Colliding with 
the target photons, a primary photon or electron produce 
e-m cascade due to reactions $\gamma+\gamma_{\rm tar} \to e^++e^-$,
$e+\gamma_{\rm tar} \to e'+\gamma'$, etc (see Fig.~\ref{cascade}).
The standard case is given
by production of HE neutrinos in extragalactic space, where cascade 
develops due to collisions with CMB photons ($\gamma_{\rm tar}= 
\gamma_{\rm CMB}$). In case the neutrino production occurs in a galaxy,
the accompanying photon can either freely escapes from a galaxy
and produce cascade in extragalactic space, or produce cascade on 
the background radiation (e.g. infra-red) within the galaxy. In the 
latter case the galaxy should be transparent for the cascade photons 
in the range 10~ MeV - 100~GeV. 

The spectrum of the cascade photons is calculated \cite{BeSm,book,Blasi-cas}:
in low energy part it is $\propto E^{-3/2}$, at high energies  
$\propto E^{-2}$ with a cutoff at some energy $\epsilon_{\gamma}$.
The energy of transition between two regimes is given approximately by
$\epsilon_c \approx (\epsilon_t /3)(\epsilon_{\gamma} /m_e)^2$, 
where $\epsilon_t$ is the mean energy of the target photon. 
In case the cascade develops in extragalactic space
$\epsilon_t=6.35\times 10^{-4}$~eV,
$\epsilon_{\gamma} \sim 100$~GeV (absorption on optical radiation),
and $\epsilon_c \sim 8$~MeV. The cascade spectrum is very close 
to the EGRET observations in the range 3~MeV - 100~GeV \cite{EGRET}.  
The observed energy density in this range is 
$\omega_{\rm EGRET} \approx (2 - 3)\times 10^{-6}$~eV/cm$^3$. 
The upper limit on HE neutrino flux $J_{\nu}(>E)$ 
is given by chain of the following inequalities  
$$
\omega_{\rm cas}>\frac{4\pi}{c}\int_E^{\infty}EJ_{\nu}(E)dE>
\frac{4\pi}{c}E\int_E^{\infty}J_{\nu}(E)dE\equiv \frac{4\pi}{c}EJ_{\nu}(>E),
$$
which in terms of the differential neutrino spectrum $J_{\nu}(E)$ 
results in
\begin{equation}
E^2 J_{\nu}(E) < \frac{c}{4\pi}\omega_{\rm cas },~~ {\rm with}~
\omega_{\rm cas} <\omega_{\rm EGRET}.
\label{cas-rig}
\end{equation}
Unless otherwise is stated, here and everywhere below the neutrino 
flux $J_{\nu}$ is given as sum of all neutrino flavors.   

Eq. (\ref{cas-rig}) gives the {\em rigorous} upper limit on the neutrino flux. 
It is valid for neutrino production
by HE protons, by TDs, by
annihilation and decays of superheavy particles, i.e. in all cases
when neutrinos are produced through decay of pions and kaons. It is
valid for production of neutrinos in extragalactic space and in
galaxies, if they are transparent for the cascade photons. It holds
for arbitrary neutrino spectrum falling down with energy. If one assumes 
some specific shape of neutrino spectrum, the cascade limit becomes stronger.   
For example, for $E^{-2}$ neutrino spectrum one immediately obtains 
\begin{equation}
E^2J_{\nu}(E) \leq \frac{c}{4\pi}\frac{\omega_{\rm cas}}
{\ln (E_{\rm max}/E_{\rm min})},
\label{cas-E2}
\end{equation}

%%%%%%%%%%%%%%%%%%%%%%%%%%%%%%%%%%%%%%%%%%%%%%%%%%%%%%%%%%%%%%%%%%%%%%%%%%%%
\section{Cosmogenic neutrinos }
Cosmogenic has a meaning ``produced by cosmic rays''. 

The most efficient mechanism for production of cosmogenic superGZK neutrinos
is given by $p\gamma$ collisions of protons with CMB photons: for  
$E_{\nu} \gsim 1\times 10^{20}$~eV the energy of the parent protons 
$E_p \sim 20 E_{\nu}$ is enough for photopion production in collisions 
with CMB photons. The space density of CMB photons (412 cm$^{-3}$)
is usually much larger than number density of the gas  and optical/IR photons 
in the sources and outside. 

We shall reproduce here the historically first calculations \cite{BZ} of 
the diffuse neutrino flux produced by UHE protons colliding with CMB
photons. 

Consider the universe filled uniformly by UHECR sources with space
density $n_s$ and UHE proton luminosity $L_p$. We assume the
cosmological evolution of the sources in the form 
${\cal L}={\cal L}_0 (1+z)^m$, where ${\cal L}_0=L_p n_s$ is
emissivity at the epoch with redshift $z=0$ and factor $(1+z)^m$ describes 
the evolution.

The production rate of a source is given as
\begin{equation}
Q_{\rm gen}=(\gamma_g -2)L_p E^{-\gamma_g},    
\label{rate}
\end{equation}
where all energies are measured in GeV and $E_{\rm min}$ is assumed to
be $\sim 1$~GeV. 

The diffuse UHE proton flux can be calculated from particle conservation, 
assuming the generation energy $E_g=E_g(E,z)$ due to energy losses and 
integrating over all epochs of generation:
\begin{equation}
J_p(E)=\frac{c}{4\pi}(\gamma_g - 2){\cal L}_0\int dt (1+z)^m 
E_g^{-\gamma_g}(E,z)dE_g/dE.
\label{diff}
\end{equation}

{\it Unmodified diffuse spectrum} is calculated with
only adiabatic energy losses included: $E_g(E,z)=(1+z)E$ and 
$dE_g/dE=1+z$. Using the connection of cosmological time $t$ and redshift
$z$ as $dz/dt= H_0 (1+z) \lambda(z)$, where
\begin{equation}
\lambda(z)=\sqrt{(1+z)^3\Omega_m + (1+z)^2 \Omega_r+\Omega_{\Lambda}},
\label{lambda}
\end{equation}
$H_0$ is the Hubble constant and $\Omega_m$,~ $\Omega_r$ and  
$\Omega_{\lambda}$ are cosmological density in units of critical
density of non-relativistic dark matter $m$, relativistic dark matter $r$ and
that due to vacuum energy density $\Lambda$, respectively, one obtains for the 
unmodified spectrum of UHE protons:
\begin{equation}
J_{\rm unm}(E)=\frac{c}{4\pi}(\gamma_g - 2)\frac{{\cal L}_0}{H_0}
E^{-\gamma_g} \eta_{\rm ev}(m,z_{\rm max}) ,
\label{unm}
\end{equation}
where $\eta_{\rm ev}(m,z_{\rm max},\gamma_g)$ is the evolutionary factor
given by
\be
\eta_{\rm ev}(m,z_{\rm max},\gamma_g)=\int_0^{z_{\rm max}} 
dz (1+z)^{m-\gamma_g}/\lambda(z).
\label{evol-factor}
\ee
It is easy to express the neutrino diffuse flux through unmodified
proton flux, assuming that a proton undergoes several collisions with 
CMB photons:
\begin{equation}
J_{\nu}(E)=\frac{2}{3}\cdot 3 \left
(\frac{E_{\nu}}{E_p}\right )^{\gamma_g-1}\frac{1}{1-\alpha^{\gamma_g-1}}
J_{\rm unm}(E),
\label{nu-flux}
\end{equation}
where 2/3 accounts for probability of charged pion production, 3 -  for 3  
neutrinos produced in the chain of pion decay,  
$E_{\nu}/E_p \approx 0.05$ is a fraction of proton energy transferred
to neutrino, and $\alpha$ is a fraction of energy lost by the proton
in $p\gamma$ collision ($\alpha$ varies from 0.22 in
$\Delta$-resonance to 0.5 at extremely high energies); the term with 
$\alpha$ in Eq.~(\ref{nu-flux})
describes approximately the subsequent $p\gamma$ collisions of UHE proton.

The low-energy edge of neutrino spectrum (\ref{nu-flux}) is determined
by energy of a proton at epoch $z$, at which the energy loss due to pion
production becomes less than that due $e^+e^-$-pair production. 

Neutrino flux given by Eq.~({\ref{nu-flux}) strongly depends on the
parameters of cosmological evolution of the sources, $m$ and $z_{\rm max}$, 
as it is seen from Eqs.~(\ref{unm}) and (\ref{evol-factor}).

The accuracy of neutrino-flux calculations by method
of Ref.~\cite{BZ} can be compared with exact calculations of Ref.~\cite{BeGa}, 
where all details of $p\gamma$ interaction were included and fluxes were
computed for all neutrino flavors separately. The total neutrino moments
(yields) $Z_{p\gamma}(\gamma_g)$ calculated in  Ref.~\cite{BeGa} must
coincide with the coefficient in front of $J_{\rm unm}(E)$ in the rhs 
of Eq.~(\ref{nu-flux}). For $\gamma_g \leq 2.7$ the agreement is
indeed  20 - 40 \%. 

%We shall compare the accuracy of neutrino flux calculations by method
%of Ref.~\cite{BZ} with the calculations of Ref.~\cite{BeGa}, where all 
%details of $p\gamma$ interaction were included and fluxes were
%computed for all neutrino flavors separately. The neutrino flux in
%these calculations can be presented as 
%\be
%J_{\nu}(E)=Z_{p\gamma}(\gamma_g)J_{\rm unm}(E),
%\ee
%where $Z_{p\gamma}(\gamma_g)$ are total moments calculated in \cite{BeGa}.  
%Comparison shows that Eq.~(\ref{nu-flux}) has an accuracy 20 - 40 \% 
%for $\gamma_g \leq 2.7$.

The detailed calculations of UHE neutrino fluxes in the similar models 
with evolution of the sources and with normalization of the fluxes by 
the proton component have been performed in Refs. \cite{ESS} - \cite{Fod}.
In this paper I will present fluxes calculated \cite{BGG-nu} in the 
BGG models \cite{BGG}, which describe precisely the observed UHECR 
spectra, with the dip at $ 1\times 10^{18} - 4\times 10^{19}$~eV as 
the most prominent feature. 

In fact there are several BGG \cite{BGG} models which give good
fit to UHECR spectra as measured by AGASA, HiRes, Fly's Eye
and Yakutsk detectors. They differ by cosmological evolution  
of the sources, described by factor $(1+z)^m$, by exponent  $\gamma_g$
of the generation spectrum and by maximum energy of accelerated
protons $E_{\rm max}$.  
The model with the best fit (see \cite{BGG-PLB}) 
corresponds to $m=0$ (the non-evolutionary model), 
with the generation index $\gamma_g=2.7$ and  
$E_{\rm max}=1\times 10^{21}$~eV. Less important assumption is
flattening of generation spectrum at $E \leq E_c$, with 
$E_c \sim 1\times 10^{18}$~eV, which is needed to describe correctly
the mass composition observed at $E \leq 1\times 10^{17}$~eV \cite{BGH}.  
 \begin{figure*}[h]
  \begin{center}
  \mbox{\includegraphics[width=12cm]{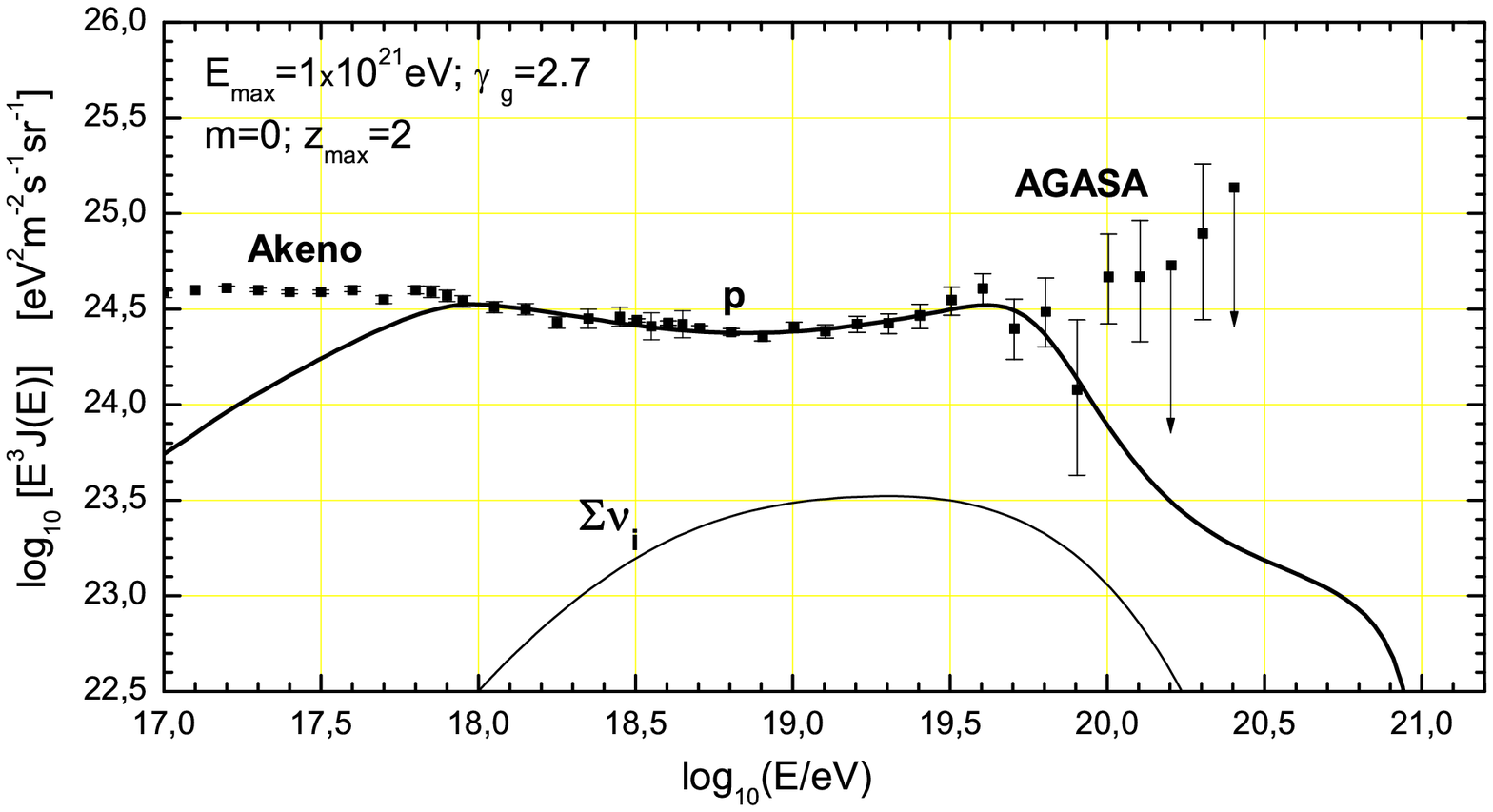}}
  \end{center}
\begin{center}
  \mbox{\includegraphics[width=12cm]{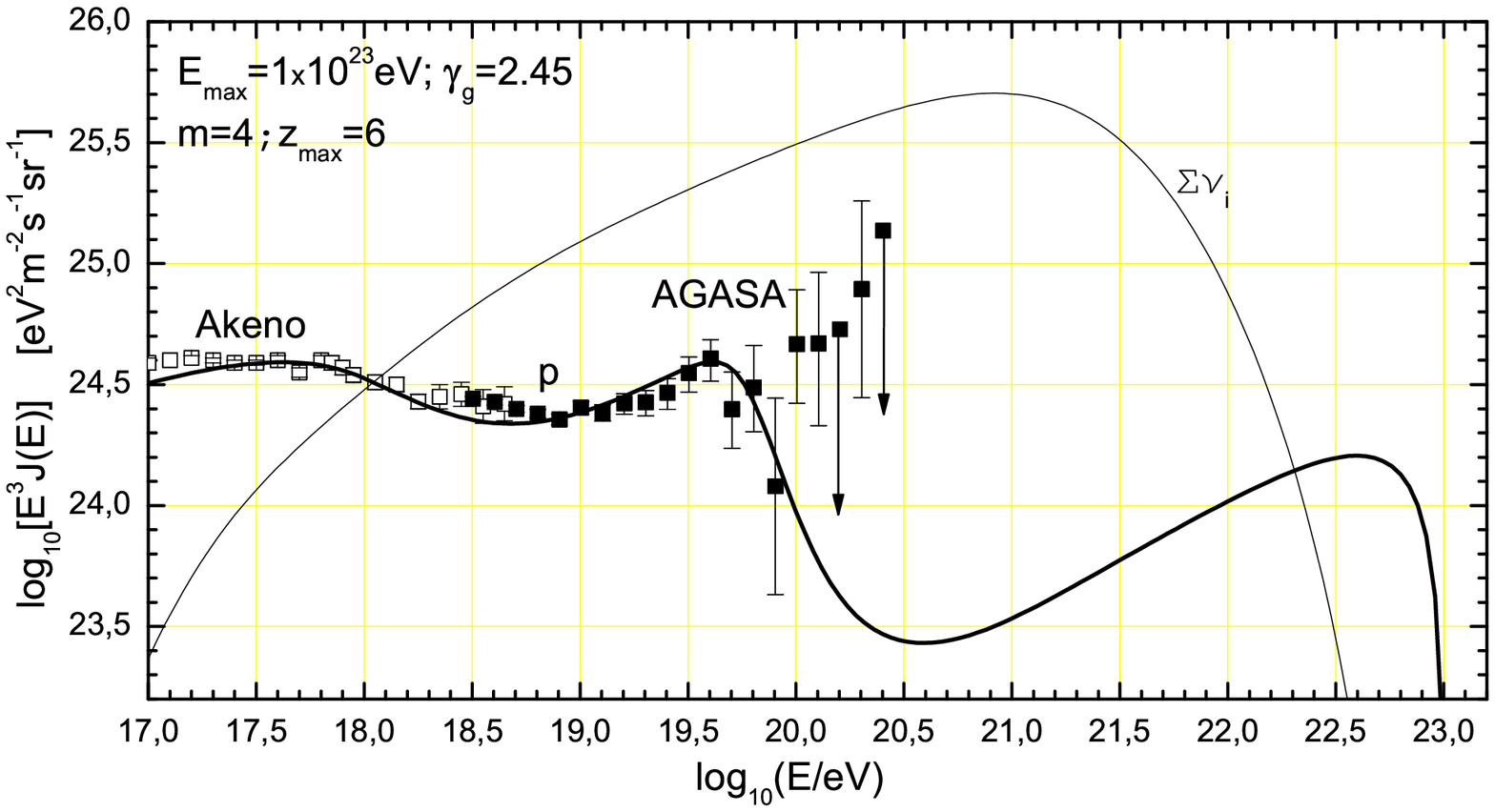}}
  \end{center}
\vspace*{-2mm}
\caption{UHE neutrino fluxes \protect\cite{BGG-nu} in the non-evolutionary 
 (upper panel) and 
 evolutionary (lower panel)  BGG models \protect\cite{BGG}.
 In the upper panel the neutrino flux accompanying the
 observed UHECR flux is shown by curve $\Sigma\nu_i$. The following 
  parameters are used in calculations: $m=0$,~ $z_{\rm max}=2$,~ 
  $\gamma_g=2.7$,~
  $E_{\rm max}=1\times 10^{21}$~eV,~ $E_c=1\times 10^{18}$~eV and
  emissivity ${\cal L}_0=3.5\times 10^{46}$~erg/Mpc$^3$yr.  
  In the low panel the neutrino flux is maximized by the following
  choice of parameters: $m=4.0$, $z_{\rm max}=6.0$,~ $\gamma_g=2.45$ 
  and emissivity ${\cal L}_0= 1.2\times 10^{46}$~erg Mpc$^{-1}$yr$^{-1}$.
  Note that fit to observational data in case of the evolutionary 
  model is worse than for non-evolutionary model.
 }
 \label{uhenu-BGG}
\end{figure*} 
The spectrum of CR in this model is shown in Fig.~\ref{uhenu-BGG} 
(upper panel) in comparison with Akeno-AGASA data. 
The emissivity of the sources needed
to fit the the observed flux is 
${\cal L}_0=3.5\times 10^{46}$~erg/Mpc$^3$yr, which corresponds to
luminosity of a source $L_p=3.7 \times 10^{43}$~erg/s for space density of
the sources (powerful AGN) $n_s=3\times 10^{-5}$~Mpc$^{-3}$.
One can notice the 
precise agreement with the Akeno-AGASA data at $1\times 10^{18} 
- 8\times 10^{19}$~eV. The explanation of the AGASA excess at 
$E> 1\times 10^{20}$~eV 
needs the additional CR component of another origin. 
The calculated neutrino
flux is shown by curve $\Sigma \nu_i$ for sum of all neutrino flavor.
{\it This is the lowest neutrino flux compatible with the observed UHECR   
flux}, because including evolution and increasing $E_{\rm max}$
one increases the neutrino flux. The predicted flux of superGZK neutrinos 
at $E \gsim 1\times 10^{20}$~eV is hardly detectable by the methods
discussed above. 

{\it Thus, the observed UHECR flux does not guarantee the detectable flux of 
superGZK neutrinos.} 

In Fig.~\ref{uhenu-BGG} (lower panel) neutrino flux is maximized for the BGG
models, using the cosmological evolution, compatible with the observed 
UHECR flux, namely ${\cal L}(z)=(1+z)^m{\cal L}_0$ at $z \leq z_{\rm max}$  
with $m=4$, $z_{\rm max}=6$ and ${\cal L}_0=1.2\times 10^{46}$~erg/Mpc$^3$yr.
Very large $E_{\rm max}=1\times 10^{23}$~eV is used to
increase further neutrino flux. Note, that though the local emissivity 
at $z=0$ is relatively low, it was much larger in
the past. The maximum acceleration energy  $E_{\rm max}$
cannot be also considered as realistic, and used mostly for
illustration of range in the predicted fluxes. Finally, this model 
fits observed UHECR spectrum worse than non-evolutionary model and has 
the problems with observed mass composition of cosmic rays at 
$E < 1\times 10^{18}$~eV.  

In the evolutionary models with large $m$ and $z_{\rm max}$, 
which explain the observed UHECR, the fluxes
of cosmogenic neutrinos are observable  by future detectors. 
\begin{figure*}[htb]
  \begin{center}
  \mbox{\includegraphics[width=0.7\textwidth]{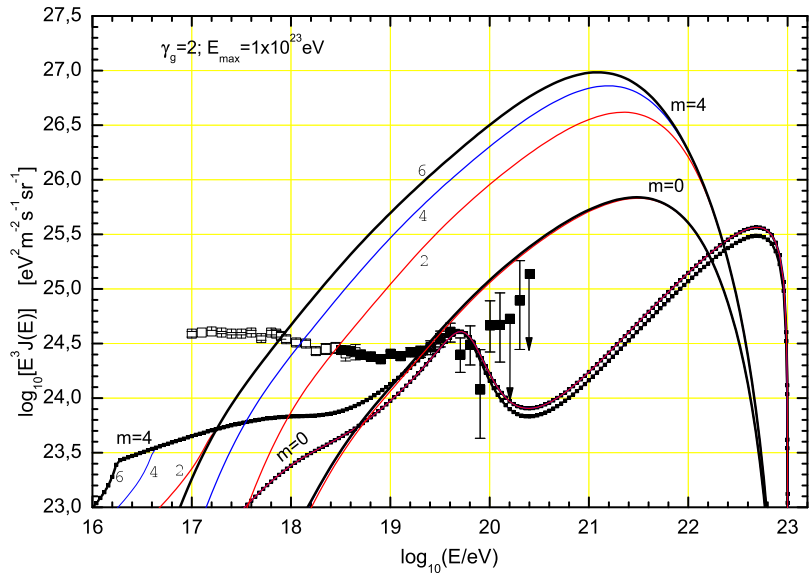}}
  \end{center}
  \caption{ UHE neutrino fluxes \protect\cite{BGG-nu} in the evolutionary 
  and non-evolutionary models with proton generation spectrum  
  $\propto E^{-2}$ and maximum acceleration energy  
  $E_{\rm max}=1\times 10^{23}$~eV. The calculated proton spectra 
  (solid curves with dots) for $m=0$ and $m=4$  are shown in
  comparison with the Akeno-AGASA data. The figures on spectrum curves
  ( 2,~ 4 and 6) show $z_{\rm max}$. The emissivities are 
  ${\cal L}_0= 2.65\times 10^{45} $~erg Mpc$^{-1}$yr$^{-1}$ for $m=0$
  and ${\cal L}_0= 2.2\times 10^{45} $~erg Mpc$^{-1}$yr$^{-1}$
  for $m=4$.
  }
\label{uhenuE2}
  \end{figure*}

The largest cosmogenic neutrino flux can be obtained in the models
with flat proton generation spectra. This class of models cannot
describe the spectrum at energies $1\times 10^{18} - 1\times 10^{19}$~eV  
and thus corresponds to transition from galactic to extragalactic 
cosmic rays at the ankle $E \sim 1\times 10^{19}$~eV. It has been
proposed \cite{SS} that large neutrino flux can be a signature of the 
ankle as transition from galactic to extragalactic cosmic rays. 
The calculated
neutrino spectra for the proton generation spectrum $\propto E^{-2}$ 
and $E_{\rm max} =1\times 10^{23}$~eV is presented in Fig.~\ref{uhenuE2}
for non-evolutionary model $m=0$ and for the evolutionary models with 
with $m=4$ and $z_{\rm max}= 2,~ 4$ and 6. The emissivity ${\cal L}_0$ 
at $z=0$ varies from $2.2\times 10^{45}$~erg Mpc$^{-3}$ yr$^{-1}$ for 
$m=4$ to $2.7\times 10^{45}$~erg Mpc$^{-3}$ yr$^{-1}$ for $m=0$.  

The largest neutrino flux in  Fig.~\ref{uhenuE2}
($m=4$ and $z_{\rm max}=6$) almost saturates the
cascade upper limit: $\omega_{\rm cas}=1.5\times 10^{-6}$~eV/cm$^{3}$, 
to be compared with the EGRET upper limit 
$\omega_{\rm cas} \approx 2\times 10^{-6}$~ eV/cm$^{3}$.

Fig.~\ref{uhenuE2} illustrates the range of predictions for UHE
neutrino fluxes in case of flat generation spectrum: from modest flux 
in the non-evolutionary case $m=0$ to flux 30 times higher in case of
evolution.  However, in all cases the maximum acceleration energy
$E_{\rm max}=1\times 10^{23}$~eV is at least two orders above that 
obtained in realistic models. The top-down scenarios considered in the
next sections provide  these high energies naturally.

%%%%%%%%%%%%%%%%%%%%%%%%%%%%%%%%%%%%%%%%%%%%%%%%%%%%%%%%%%%%%%%%%%%
\section{UHE neutrinos from Superheavy Dark Matter (SHDM)}  
SHDM is one of the models for cosmological cold dark matter 
\cite{BKV,whimpzillas}. The most attractive mechanism of
production is given by creation of superheavy particles in
time-varying gravitational field in post-inflation 
epoch \cite{Kolb-grav,Kuz-grav}. Creation 
occurs when the Hubble parameter is of order of particle mass 
$H(t) \sim m_X$. Since the maximum value of the Hubble parameter is limited  
by the mass of the inflaton $H(t) \lsim m_{\phi} \sim 10^{13}$~GeV,  
the mass of X-particle is limited by $m_{\phi}$, too. For example, 
$m_X \sim 3\times 10^{13}$~GeV results in $\Omega_X h^2 \sim 0.1$, as
required by WMAP measurements. 

Being protected by some symmetry, SHDM particles with such masses can be 
stable or quasi-stable. In case of gauge symmetry they are stable, in
case of gauge discrete symmetry they can be stable or quasi-stable. 
Decay can be provided by superweak effects: wormholes, instantons, 
high-dimension operators etc. 

Like any other form of cold dark matter, X-particles are accumulated
in the halo with overdensity $2.1\times 10^{5}$. 

SHDM particles can produce UHECR and high energy neutrinos at the
decay of X-particles (when the protecting symmetry is broken) and at
their annihilation, when the symmetry is exact. The scenario with 
decaying X-particles was first studied in \cite{BKV,KR,Sarkar}.
An interesting scenario with stable X-particles, when UHE particles
are produced by annihilation of X-particles has been put forward in 
\cite{Khlopov}. 
In this scenario superheavy X-particles have the gauge charge and they
are produced at post-inflationary epoch by close pairs, forming the 
bound systems. Loosing the
angular momentum, these particles inevitably annihilate in a close
pair.   

The UHE particles (protons, pions and neutrinos from 
the chain of pion decays)  are produced as a result of QCD cascading
of partons. The calculations of fluxes and spectra are nowadays
reliably performed by Monte Carlo \cite{BK-MC} and using the  
DGLAP equations \cite{DGLAP-Rub} - \cite{DGLAP-ABK}.  
The spectra of protons, photons and 
neutrinos are shown in Fig.~\ref{SHDM} for the case of SHDM particles 
with mass $M_X=1\times 10^{14}$~GeV. One can observe the large fluxes
of superGZK neutrinos with very high energies in excess of $10^{22}$~eV.  
\begin{figure*}[h]
  \begin{center}
  \mbox{\includegraphics[width=0.7\textwidth]{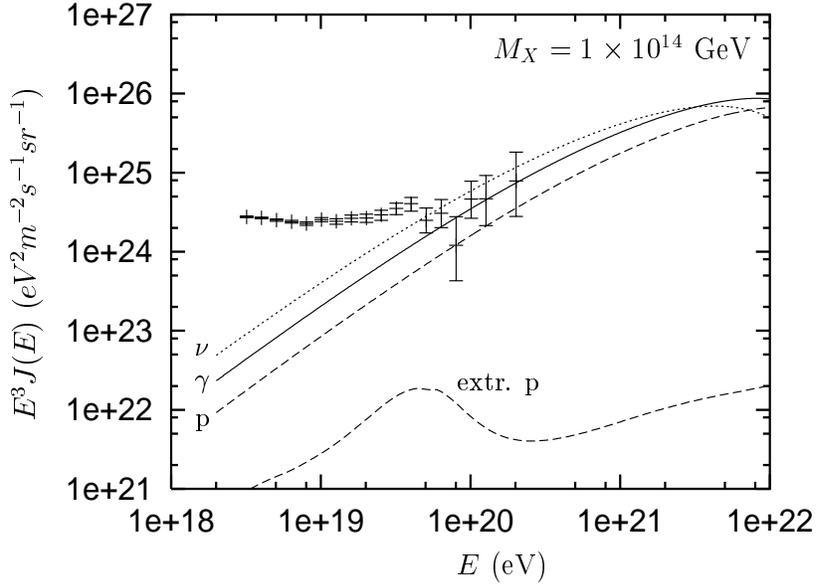}}
  \end{center}
  \caption{Spectra of neutrinos (upper curve), photons (middle curve)
    and protons (two lower curves) in SHDM model compared with AGASA
    data, according to calculations of \protect\cite{DGLAP-ABK}.
    The neutrino flux is dominated by the halo
    component with small admixture of extragalactic flux. The flux of
    extragalactic protons is shown by the lower curve (extr. p).}
    \label{SHDM}
  \end{figure*}

\section{SuperGZK neutrinos from Topological Defects (TDs)}
As has been first noticed by D.~A.~Kirzhnitz \cite{Kirzhnitz},
each spontaneous symmetry breaking in the early universe is accompanied by 
the phase transition. Like the phase transitions in liquids and
solids, the cosmological phase transitions can give rise to
topological defects (TDs), which can be in the form of surfaces
(cosmic textures), lines (cosmic strings)
and points (monopoles). 
In many cases TDs become unstable and decompose to constituent fields, 
superheavy gauge and Higgs bosons (X-particles), which then decay 
producing UHECR. It could happen, for example, when two segments of 
ordinary string, or monopole and antimonopole touch each other, when 
electrical current in superconducting string reaches the critical value
and in some other cases. The decays of these particles, if they heavy enough, 
produce particles of ultrahigh energies including neutrinos.

The following TDs are of interest for UHECR and neutrinos:\\  
{\em monopoles} ($G \to H\times U(1)$ symmetry breaking), 
{\em ordinary strings}
($U(1)$ symmetry breaking) with important subclass of 
superconducting strings,  {\em monopoles connected by strings} 
($G \to H\times U(1)$ symmetry breaking with subsequent $U(1) \to Z_N$
symmetry breaking, where $Z_N$ is discrete symmetry). The important
subclass of the monopole-string network is given by {\em necklaces},
when $Z_N=Z_2$, i.e. each monopole is attached to two strings. 
We shall shortly describe the production of UHE particles by these TDs.

(i) {\em Superconducting strings}.\\
As was first noted by Witten\cite{Witten}, in a wide class of elementary 
particle models, strings behave like superconducting wires. Moving through 
cosmic magnetic fields, such strings develop electric currents.
Superconducting strings produce X particles when the electric current
in the strings reaches the critical value. Superconducting strings
produce too small flux of UHE particles \cite{BBV} to be the sources 
of observed UHECR.

(ii) {\em Ordinary strings}.\\
There are several mechanisms by which ordinary strings can produce UHE 
particles.

For a special choice of initial conditions, an ordinary  string loop
can collapse to a
double line, releasing its total energy in the form of X-particles. 
However, the probability of this mode of collapse is
extremely small, and its contribution to the overall flux of UHE
particles is negligible.

String loops can also 
produce X-particles when they self-intersect.
Each intersection, however, gives only a few
particles, and the corresponding flux is very small. 

Superheavy particles with large Lorentz factors can be produced in 
the annihilation of cusps, when the two cusp segments overlap.  
The energy released in a single cusp event can be quite large, but
again, the resulting flux of UHE particles is too small to account for
the observations.

It has been argued \cite{Vincent} that long
strings lose most of
their energy not by production of closed loops, as it is generally
believed, but by direct emission of heavy X-particles.
If correct, this claim will change dramatically 
the standard picture of string evolution. It has been also
suggested that the decay products of particles produced in this
way can explain the observed flux of UHECR \cite{Vincent}. 
However, as it is argued in Ref.~\cite{BBV}, numerical simulations described in
\cite{Vincent} allow an alternative interpretation not connected with 
UHE particle production.

(iii){\em Network of  monopoles connected by strings}.\\
The sequence of phase transitions
\begin{equation}
G\to H\times U(1)\to H\times Z_N
\label{symm}
\end{equation}
 results in the formation of monopole-string networks in which each monopole 
is attached to N strings. Most of the monopoles and most of the strings belong 
to one infinite network. The evolution of networks is expected to be 
scale-invariant with a characteristic distance between monopoles 
$d=\kappa t$, where $t$ is the age of Universe and $\kappa=const$. 
The production of UHE particles are considered in \cite{BMV}. Each 
string attached 
to a monopole pulls it with a force equal to the string tension, $\mu \sim 
\eta_s^2$, where $\eta_s$ is the symmetry breaking vev of strings. Then
monopoles have a typical acceleration $a\sim \mu/m$, energy $E \sim \mu d$ 
and Lorentz factor $\Gamma_m \sim \mu d/m $, where $m$ is the mass of the 
monopole. Monopole moving with acceleration can, in principle, radiate  
gauge quanta, such as photons, gluons and weak gauge bosons, if the
mass of gauge quantum (or the virtuality $Q^2$ in the case of gluon) is
smaller than the monopole acceleration. The typical energy of radiated quanta 
in this case is $\epsilon \sim \Gamma_m a$. This energy can be much higher 
than what 
is observed in UHECR. However, the produced flux (see \cite{BBV}) is much 
smaller than the observed one. 

(vi){\em Necklaces}.\\
Necklaces are hybrid TDs corresponding to the case $N=2$ , i.e. to the
case when each monopole is attached to two
strings.  This system resembles ``ordinary'' cosmic strings,
except the strings look like necklaces with monopoles playing the role
of beads. The evolution of necklaces depends strongly on the parameter
\begin{equation}
r=m/\mu d,
\end{equation}
where $m$ is a mass of a monopole, $\mu$ is mass per unit length of a 
string (tension of a string) and 
$d$ is the average separation between monopoles and antimonopoles
along  the strings.
As it is argued in Ref.~\cite{BV}, necklaces might evolve to  
configurations with $r\gg 1$.  
Monopoles and antimonopoles trapped in the necklaces
inevitably  annihilate in the end, producing first the heavy  Higgs and 
gauge bosons ($X$-particles) and then hadrons.
The rate of $X$-particle production can be estimated as \cite{BV} 
\begin{equation}
\dot{n}_X \sim \frac{r^2\mu}{t^3m_X}.
\label{xrate}
\end{equation}
This rate determines the rates of pion and neutrino production
with energy spectrum calculated in Ref.~\cite{DGLAP-ABK}. 

Restriction due to e-m cascade radiation demands the cascade energy density 
$\omega_{cas} \leq 2\cdot 10^{-6}$~eV/cm$^3$. The cascade energy density 
produced by necklaces can be calculated as
\begin{equation}
\omega_{cas}=
\frac{1}{2}f_{\pi}r^2\mu \int_0 ^{t_0}\frac{dt}{t^3}
\frac{1}{(1+z)^4}=\frac{3}{4}f_{\pi}r^2\frac{\mu}{t_0^2},
\label{eq:n-cas}
\end{equation}
where $f_{\pi}\approx 0.5$ is a fraction of total energy release 
transferred to the cascade. Therefore, $r^2\mu$ 
and the rate of X-particle production (\ref{xrate}) is limited by cascade 
radiation.

The fluxes of UHE protons, photons and neutrinos from  are shown in 
Fig.~\ref{neckl} according to 
calculations of \cite{DGLAP-ABK}. The mass of X-particle
is taken $m_X=1\times 10^{14}$~GeV. Neutrino flux is noticeably higher
than in the case of conservative scenarios for  cosmogenic neutrinos
and neutrinos from SHDM.
\begin{figure*}[h]
  \begin{center}
  \mbox{\includegraphics[width=0.7\textwidth]{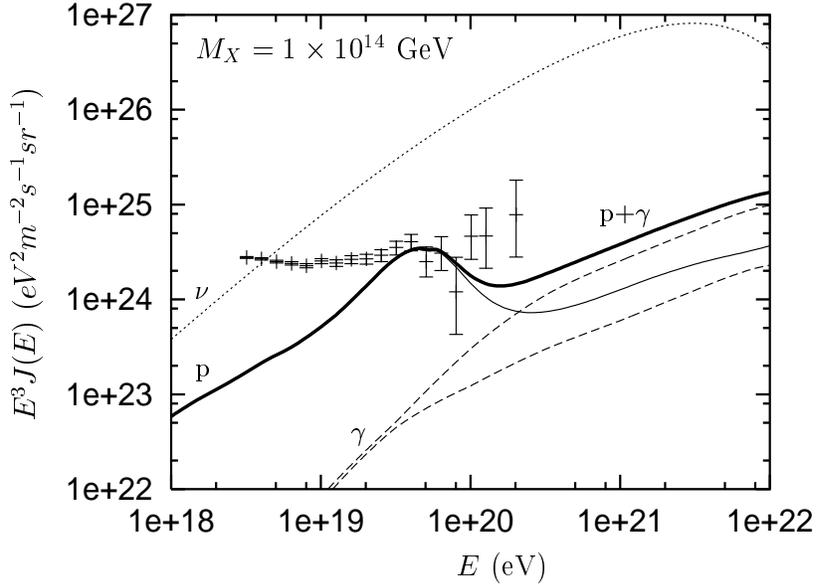}}
  \end{center}
  \caption{Diffuse spectra of neutrinos, protons and photons from
  necklaces. The upper curve shows neutrino flux, the middle - proton
  flux and two lower curves - photon fluxes for two cases of
  absorption. The thick curve gives the sum of the proton and the higher 
  photon flux.}
  \label{neckl}
  \end{figure*}
\section{Mirror neutrinos}
Mirror matter can be most powerful source of superGZK neutrinos not limited
by the usual cascade limit \cite{mirror}. 

Existence of mirror matter is based on the deep theoretical concept, which
was  introduced by Lee and Yang \cite{LY}, Landau \cite{Landau} and 
most notably
by Kobzarev, Okun and Pomeranchuk \cite{KOP}. Particle space is a
representation of the Poincare group. Since the space reflection 
$\vec{x} \to -\vec{x}$ and
time shift $t \to t+\Delta t$ 
commute as the coordinate transformations, the
corresponding inversion operator $I_s$ and the Hamiltonian $H$ must
commute, too: $[I_s,H]=0$. Because the parity operator $P$ does not commute 
with $H$ (i.e. parity is not conserved) Lee and Yang suggested that 
$I_s=P\cdot R$, where the operator $R$ generates the mirror particle 
space, and thus $I_s$ transfers 
the left states of ordinary particles into right states of the mirror 
particles and vise versa. In fact, the assumption of Landau is
similar: one may say that he assumed $R=C$. 

The mirror particles have interactions identical 
to the ordinary particles, but these two sectors interact with each other
only gravitationally \cite{KOP}. Gravitational interaction mixes the visible 
and mirror neutrino states, and thus causes the oscillation between
them.  
\begin{figure*}[htb]
\begin{center}
\mbox{\includegraphics[width=0.7\textwidth]{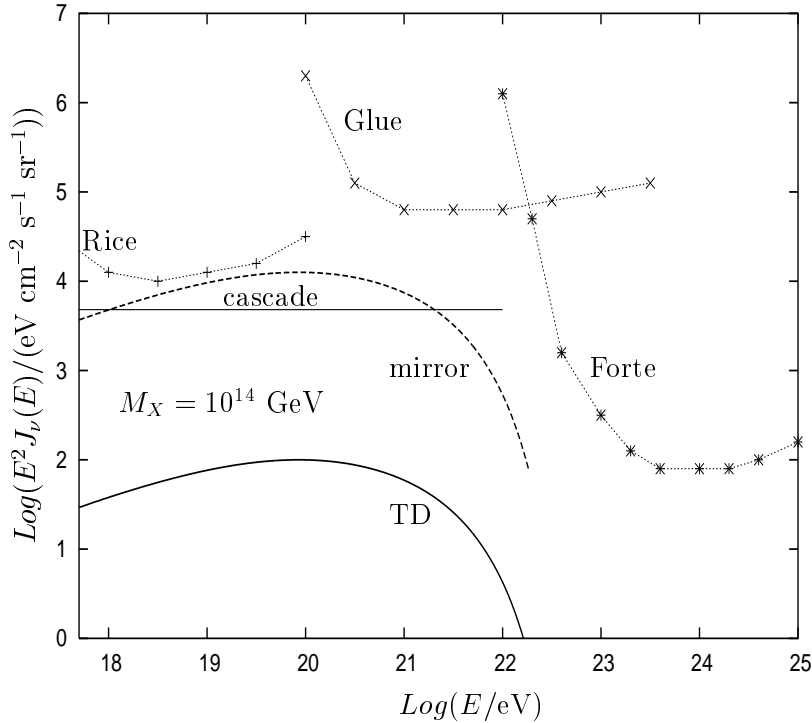}}
\end{center}
\caption{
Diffuse flux of the visible neutrinos from mirror necklaces 
with $M_X=1\times 10^{14}$~GeV. The flux is limited by observations 
of RICE, GLUE and FORTE. Note, that neutrino flux exceeds the
general cascade upper limit (\ref{cas-rig}). TD curve gives the flux
from the ordinary necklaces.
}
\label{mirr-nu}
\end{figure*}

A cosmological scenario must provide the suppression of the mirror
matter and in particular the density of mirror photons and neutrinos 
at the epoch 
of nucleosynthesis. It can be obtained in the two-inflaton model 
\cite{mirror}. The rolling of two inlatons to minimum of the potential
is not synchronized, and when the mirror inflaton reaches minimum, the ordinary
inflaton continues its rolling, inflating thus the mirror matter produced
by the mirror inflaton. While mirror matter density is suppressed, the mirror
topological defects can strongly dominate \cite{mirror}. Mirror TDs 
copiously produce mirror neutrinos with extremely high
energies typical for TDs, and they are not accompanied by
any visible particles. Therefore, the upper limits on HE mirror 
neutrinos in our world do not exist. All HE mirror particles 
produced by mirror TDs are sterile for us, interacting with
ordinary matter only gravitationally, and only mirror neutrinos 
can be efficiently converted into ordinary ones due to oscillations. 
The only (weak) upper limit comes from the resonant
interaction  of converted neutrinos with DM neutrinos: 
$\nu+\bar{\nu}_{\rm DM}\to Z^0$ \cite{mirror}.
We shall obtain here this upper limit in the simplified case of 
degenerate neutrinos with common mass $m_{\nu}$.

The cascade energy density can be calculated as 
\begin{equation}
\omega_{cas}=
2\pi \frac{f_h}{f_{tot}}\sigma_t n_{\nu_i} t_0 E_0^2 I_{\nu}(E_0),
\label{omega}
\end{equation}
where 
$$
E_0=\frac{m_Z^2}{2m_{\nu}}=1.81\cdot 10^{13}\left  
(\frac{0.23~{\rm eV}}{m_{\nu}}\right )~{\rm GeV}
$$
is the resonant neutrino energy, $n_{\nu_i}$ is the density of DM
neutrinos, $f_{tot}$ and $f_{had}$ are total and hadron widths of 
$Z^0$ decay, respectively, and 
\begin{equation}
\sigma_t = 48 \pi f_{\nu}G_F=1.29\cdot 10^{-32}~{\rm cm}^2,
\label{eff}
\end{equation}
is the effective $\nu\bar{\nu}$-cross-section in the resonance. \\
Eq.~(\ref{omega}) gives the upper bound on $I_{\nu}(E_0)$ which is
very weak, due to factor $\sigma_t n_{\nu_i} t_0$, 
as compared with that for visible neutrinos. 

The strongest limit on the fluxes of superGZK neutrinos are given 
nowadays by radio observations \cite{glue} - \cite{rice}. 

The mirror neutrino flux can be calculated for the case of mirror
necklaces identically to the calculations in Section~5 for ordinary
necklaces, but with parameter $r^2/mu$ not being limited any more by 
the cascade upper bound. The probability of oscillation is given 
by $P_{\rm osc}=1/2$, since oscillation lengths are very small in
comparison with the distances to TDs. The calculated neutrino fluxes
for $M_X=1\times 10^{14}$~GeV are shown in  Fig.~\ref{mirr-nu}
together with radio upper limits. The calculated flux exceeds the
cascade upper limit for ordinary neutrino sources shown in 
Fig.~\ref{mirr-nu}.

%%%%%%%%%%%%%%%%%%%%%%%%%%%%%%%%%%%%%%%%%%%%%%%%%%%%%%%%%%%%%%%%%%%%%
\section{Conclusions}
SuperGZK neutrinos with energies higher than $1\times 10^{20}$~eV can 
be efficiently searched for by future space detectors EUSO and OWL,
and by radio methods. The neutrino-induced inclined EAS can be
detected by Auger. The future detectors can control very large area  
(up to $\sim 10^5$~km$^{2}$ in case of EUSO) and thus they are
sensitive to very low superGZK neutrino fluxes. The energy threshold 
of these methods is typically high, and it makes the superGZK
neutrinos the main goal of the search. 

The most conservative mechanism of superGZK neutrino production is
given by interaction of UHECR with CMB photons. One might think that 
the basic elements for UHE neutrino generation are reliably known: 
the beam of observed UHECR and the target, build by CMB photons. 
However, the observed flux of UHECR does not guarantee the detectable 
flux of superGZK neutrinos. A very reasonable model, which describes  
perfectly well the observed UHECR spectrum, predicts the neutrino flux 
an order of magnitude lower that of the observed UHECR flux (see 
upper panel of Fig.~\ref{uhenu-BGG}). The detectable fluxes of superGZK
neutrinos require three conditions: {\em (i)} the maximum acceleration energy 
$E_{\rm max} \gg 1\times 10^{20}$~eV, {\em (ii)} the cosmological evolution of
the UHECR sources (most probably AGN) and {\em (iii)} flat generation
spectrum (e.g. $\propto E^{-2}$ favors the large neutrino flux). 
The necessary conditions {\em (i)} and {\em (ii)}
imply the unknown astrophysics. It is especially true for {\em (i)}:
there are no reliable mechanisms of acceleration with $E_{\rm max} \sim 
10^{22} - 10^{23}$~eV, though many ideas have been put forward. 
The lower panel of Fig.~\ref{uhenu-BGG} presents the superGZK neutrino fluxes 
for the extreme hypothetical assumptions: very large $E_{\rm max}$ and
strong evolution of the sources up to $z_{\rm max} = 6$. 

The top-down scenarios predict naturally very high neutrino energies 
up to $\sim 0.1 m_{\rm GUT}$, and in some cases (monopole-string
network)  up to $m_{\rm Pl}$. The fluxes of neutrinos are also
naturally high. The neutrino fluxes are rigorously constrained by 
the cascade upper limit (\ref{cas-rig}). The mirror neutrinos do not
respect this  limit, and their fluxes can be even larger 
(see Fig.~\ref{mirr-nu}).
 
The search for superGZK neutrinos is in any case is the search for a new 
physics, either for astrophysics (the new acceleration mechanisms 
and cosmological evolution of the sources, most probably AGN) 
or for topological defects, mirror topological defects and superheavy
dark matter.
\section{Acknowledgments}
 
I am grateful to my collaborators Roberto Aloisio, Askhat Gazizov 
and Svetlana Grigorieva for joint work and many useful discussions.


\begin{thebibliography}{99}

\bibitem{GZK} 
K.~Greisen, Phys. Rev. Lett. {\bf 16}, 748 (1966),
G.~T.~Zatsepin and V.~A.~Kuzmin, Pisma Zh. Experim. Theor. Phys. {\bf 4},
114 (1966).

\bibitem{BZ}
V.~S.~Berezinsky and G.~T.~Zatsepin, Phys. Lett {\bf B 28}, 423 (1969);
V.~S.~Berezinsky and G.~T.~Zatsepin, Soviet Journal of Nuclear Physics
{\bf 11}, 111 (1970).

\bibitem{Witten}
E.~Witten, Nucl. Phys. B {\bf 249}, 557 (1985).

\bibitem{HiSch}
C.~T.~Hill, D.~N.~Schramm and T.~P.~Walker, Phys. Rev. D {\bf  36},
1007 (1987).

\bibitem{BeSm}
V.~S.~Berezinsky and A.~Yu.~Smirnov, Ap.Sp.Sci {\bf 32}, 461 (1975);\\
V.~S.~Berezinsky, Proc. of ``Neutrino-77'' {\bf 1}, 177 (1977).

\bibitem{EUSO} 
see http://www.euso-misson.org/

\bibitem{OWL}
see http://heawww.gsfc.nasa.gov/docs/gamcosray/hecr/OWL/.

\bibitem{askarian}
G.~Askarian, JETP, {\bf 14} (1962) and {\bf 21} (1965). 

\bibitem{saltz}
D.~Saltzberg, Phys. Rev. Lett. {\bf 86}, 2802 (2001).

\bibitem{glue}
 P.~W.~Gorham et al, Phys. Rev. Lett. {\bf 93}, 041101 (2004)

\bibitem{forte}
N.~Lehtinen et al, Phys. Rev. D {\bf  69}, 013008 (2004)

\bibitem{rice}
I.~Kravchenko et al, astro-ph/0306408.

\bibitem{WB}
E.~Waxman and J.~Bahcall, Phys. Rev. D {\bf 59}, 023002 (1999).

\bibitem{MPR}
K.~Mannheim, R.~J.~Protheroe and J.~Rachen, Phys. Rev. D {\bf 63},
023003 (2000).

\bibitem{book}
V.~S.~Berezinsky, S.~V.~Bulanov, V.~A.~Dogiel, V.~L.~Ginzburg and
V.~S.~Ptuskin, Astrophysics of Cosmic Rays, North-Holland 1990.

\bibitem{Blasi-cas}
C.~Ferrigno, P.~Blasi, D.~De~Marco, astro-ph/0404352.

\bibitem{EGRET}
P.~Sreekumar et al. [EGRET collaboration], Astroph. J. {\bf 494}, 523 (1998).

\bibitem{BeGa}
V.~Berezinsky, A.~Gazizov, Phys. Rev. D {\bf 47}, 4206 (1993).

\bibitem{ESS}
 R.~Engel, D.~ Seckel and T.~Stanev, Phys. Rev. D {\bf 64}, 093010 (2001).

\bibitem{Kal}
O.~E.~Kalashev,  V.~A.~Kuzmin, D.~V.~Semikoz and G.~Sigl, 
Phys. Rev. D {\bf 66}, 063004 (2002).

\bibitem{Fod}
Z.~Fodor, S.~Katz, A.~Ringwald and H.~Tu, JCAP {\bf 0311}, 015 (2003).

\bibitem{BGG-nu}
V.~Berezinsky, A.~Gazizov and S.~Grigorieva in preparation.

\bibitem{BGG}
V.~Berezinsky, A.~Z.~Gazizov and S.~I.~Grigorieva, hep-ph/0204357, 
astro-ph/0210095.

\bibitem{BGG-PLB}
V.~Berezinsky, A.~Z.~Gazizov and S.~I.~Grigorieva, 
 Phys. Lett. B {\bf 612}, 147 (2005).

\bibitem{BGH}
V.~S.~Berezinsky, S.~I.~Grigorieva and B.~I.~Hnatyk,
Astropart. Phys. {\bf 21}, 617 (2004).

\bibitem{SS}
D.~Seckel, T.~Stanev, astro-ph/0502244.

\bibitem{BKV}
V.~Berezinsky, M.~Kachelriess, A.~Vilenkin, Phys. Rev. Lett. {\bf 79},
4302 (1997).

\bibitem{whimpzillas}
 E.~W.~Kolb, D.~J.~H.~Chung, A.~Riotto, Phys. Rev. Lett. {\bf 81},
 4048 (1998).

\bibitem{Kolb-grav}
 E.~W.~Kolb, D.~J.~H.~Chung, A.~Riotto, Phys. Rev. D {\bf 59},
 023501 (1999).

\bibitem{Kuz-grav}
 V.~A.~Kuzmin, I.~I.~Tkachev, JETP Lett. {\bf 68} (1998) 271-275.

\bibitem{KR}
V.~A.~Kuzmin and V.~A.~Rubakov, Phys. Atom. Nucl. {\bf 61}, 1028 (1998).

\bibitem{Sarkar}
M.~Birkel and S.~Sarkar, Astrop. Phys. , {\bf 9}, 297 (1998).

\bibitem{Khlopov}
V.~K.~Dubrovich, D.~Fargion, M.~Khlopov, Astropart. Phys. {\bf 22}, 
183 (2004).

\bibitem{BK-MC}
V.~Berezinsky, M.~Kachelriess, Phys. Rev. D {\bf 63}, 034007 (2001).

\bibitem{DGLAP-Rub}
N.~A.~Rubin, Thesis, Cavendish Laboratory University of Cambridge (1999).

\bibitem{DGLAP-Sar}
S.~Sarkar, R.~Toldra, Nucl. Phys. B {\bf 621}, 495  (2002).

\bibitem{DGLAP-Bar}
C.~Barbot, M.~Drees, Phys. Lett. B {\bf 533}, 107 (2002).

\bibitem{DGLAP-ABK}
R.~Aloisio, V.~Berezinsky, M.~Kachelriess,  Phys. Rev. D {\bf 69},
094023 (2004).

\bibitem{Kirzhnitz}
D.~A.~Kirzhnitz. JETP Lett. {\bf 15}, 745 (1975).

\bibitem{BBV}
V.~Berezinsky, P.~Blasi, A.~Vilenkin, Phys. Rev. D {\bf 58}, 103515 (1998).

\bibitem{Vincent}
G.~Vincent, N.~Antunes and M.~Hindmarsh, Phys. Rev. Lett. {\bf 80}, 2277
(1998).

\bibitem{BMV}
V.~Berezinsky, X.~Martin, A.~Vilenkin, Phys. Rev. D {\bf 56}, 2024
(1997).

\bibitem{BV}
 V.~Berezinsky, A.~Vilenkin, Phys. Rev. Lett. {\bf 79}, 5202 (1997).

\bibitem{mirror}
V.~Berezinsky and A.~Vilenkin, Phys. Rev. D {\bf 62}, 083512 (2000).

\bibitem{LY}
T.~D.~Lee and C.~N.~Yang, Phys. Rev. {\bf 104}, 254 (1956).

\bibitem{Landau}
L.~D.~Landau, JETP {\bf 32}, 405 (1957).

\bibitem{KOP}
I.~Yu.~Kobzarev, L.~B.~Okun, and I.~Ya.~Pomeranchuk, Sov. J. Nucl. Phys. 
{\bf 3}, 837 (1966).


\end{thebibliography}
\end{document}